\documentclass[showpacs,pra,twocolumn,amsmath,amssymb]{revtex4-1} %

\usepackage{epsfig}
\usepackage{subfigure}
\usepackage{graphicx}
\usepackage{dcolumn}
\usepackage{stmaryrd}
\usepackage{mathrsfs}
\usepackage{pifont}
\usepackage{amsthm}
\usepackage{bm}
\usepackage{latexsym}

\newcommand{\beq}{\begin{equation}}
\newcommand{\eeq}{\end{equation}}
\newcommand{\beqa}{\begin{eqnarray}}
\newcommand{\eeqa}{\end{eqnarray}}

\newcommand{\la}{\langle}
\newcommand{\ra}{\rangle}

\newcommand{\ti}{\tilde}
\newcommand{\ga}{\gamma}

\newcommand{\da}{\dagger}
\newcommand{\De}{\Delta}
\newcommand{\al}{\alpha}
\newcommand{\si}{\sigma}

\newcommand{\om}{\omega}

\newcommand{\non}{\nonumber}
\newcommand{\pa}{\partial}

\def\jpa#1{{ J.\ Phys.\ A} {\bf#1}}
\def\pra#1{{ Phys.\ Rev. A\/} {\bf#1}}

\def\prl#1{{ Phys.\ Rev.\ Lett.} {\bf#1}}

\def\pla#1{{ Phys.\ Lett. A\/} {\bf#1}}
\def\rmp#1{{ Rev. \ Mod. \ Phys.} {\bf#1}}

\def\epl#1{{Europhys.\ Lett. }{\bf #1}}

\begin{document}

\title{Many-Body Quantum Trajectories of Non-Markovian Open Systems}

\author{Jun Jing$^{1,4}$\footnote{Email address: Jun.Jing@stevens.edu}, Xinyu Zhao$^{1}$, J. Q. You$^{2}$, Walter T. Strunz$^{3}$, Ting Yu$^{1}$\footnote{Email address: Ting.Yu@stevens.edu}}

\affiliation{$^{1}$Center for Controlled Quantum Systems and Department of Physics and Engineering Physics, Stevens Institute of Technology, Hoboken, New Jersey 07030, USA \\ $^{2}$Laboratory for Quantum Optics and Quantum Information, Beijing Computational Science Research Center, Beijing 100084, China \\ $^{3}$Institut f\"ur Theoretische Physik, Technische Universit\"at Dresden, D-01062 Dresden, Germany\\ $^{4}$Department of Physics, Shanghai University, 200444, China}

\date{\today}

\begin{abstract}
A long-standing open problem in non-Markovian quantum state diffusion (QSD) approach to open quantum systems is to establish the non-Markovian QSD equations for multiple qubit systems. In this paper, we settle this important question by explicitly constructing a set of exact time-local QSD equations for $N$-qubit systems. Our exact time-local (convolutionless) QSD equations have paved the way towards simulating quantum dynamics of many-body open systems interacting with a common bosonic environment. The applicability of this multiple-qubit stochastic equation is exemplified by numerically solving several quantum open many-body systems concerning quantum coherence dynamics and dynamical control.
\end{abstract}

\pacs{03.65.Yz, 05.40.-a, 42.50.Lc, 37.30.+i}
\maketitle

\section{Introduction}

Dynamical aspect of quantum open systems has been investigated for many years and developed in different formalisms \cite{Gardiner_Zoller,Carmicheal,Dalibardetal,Gisin-Percival,Review,Breuer, Hu-Paz-Zhang}. Typically, the state of an open system is described by a density operator governed by a master equation that plays a pivotal role in the fields of quantum optics, quantum dissipative dynamics and quantum information. When the environment is a structured medium or the system-environment interaction is strong, Lindblad Markov master equations are prone to fail. Then non-Markovian master equations or the alternative non-Markovian approaches such as quantum trajectories or quantum jump must be used \cite{App1,App2,App3,App4,App5}. Notably, a non-Markovian quantum jump approach has been developed based on some Lindbald-type master equations \cite{nqj1,nqj2}. Despite extensive efforts, deriving an exact master equation without invoking the Born-Markov approximations (weak couplings and memoryless environment) in a fully microscopic way has achieved only limited success in practice, and is mostly restricted in a few isolated models such as the quantum Brownian motion model \cite{Hu-Paz-Zhang,Halliwell96,Strunz_Yu04}, a leaky cavity model \cite{Yu04,XZW}, a phase damping model \cite{Dephasing}, a two-level atom coupled to a quantized radiation field \cite{Atoms}, a two-body system in a collective bath \cite{EPL,Zhaoetal2011} and a single multi-level atomic system \cite{multilevel}. For a genuine many-body open system such as a multi-two-level atomic system (qubits) coupled to a fully quantized environment, the existence of an {\em exact} non-Markovian dynamical equation \cite{YDGSPLA} such as master equation is still largely unknown.

It has been shown that the non-Markovian quantum state diffusion (QSD) equations for the stochastic pure states can formally solve a quantum open system coupled to a bosonic environment irrespective of environmental memory, coupling strength and the spectral density \cite{Strunz97,Strunz98,Strunz99,Yupra99,Wisemanetal2002}. The generality of the QSD equation has given it appeal as both numerical and theoretical tools for a non-Markovian open system. On the other hand, the time-nonlocal feature of the non-Markovian QSD equations has been a major obstacle in general implementations of the trajectory formalism for a realistic physical system. Therefore, it is highly desirable to develop a time-local non-Markovian formalism that is applicable to many-body open systems.

In this paper, we report explicit constructions of exact time-local QSD equations for multiple qubit systems. In particular, we show that these exact time-local QSD equations contain only finite polynomial noise terms. Our method of deriving the exact $N$-qubit QSD equations can be modified straightforwardly to deal with other multi-atomic models with arbitrary energy-level and number of atoms. Such non-Markovian quantum trajectory equations are capable of describing the transition from non-Markovian to Markov regimes for $N$-body quantum open systems ($N\geqslant 3$).

The paper is organized as follows. In Sec.~\ref{model}, we introduce the basic idea of non-Markovian quantum trajectory approach including the linear and non-linear QSD equations. In Sec.~\ref{theorem}, a theorem is established about the explicit construction of the exact time-local QSD equation as well as the so-called O-operator for the $N$-qubit dissipative model. Then, this many-body QSD equation is employed to the quantum dynamics and non-perturbative dynamical decoupling of a three-qubit system in Sec.~\ref{result}. Finally, we conclude the paper in Sec.\ref{conclusion}. In appendices \ref{appenda} and \ref{appendb}, the details about the O-operator in $N$-qubit dissipative model and the explicit construction of the O-operator for a general three-qubit dissipative system are provided.

\section{The model and the exact QSD equation}\label{model}

A generic quantum open system in the system plus environment framework can be written as (setting $\hbar=1$):
\begin{equation}\label{Hamil}
H_{\rm tot}=H_{\rm sys}+H_{\rm int} + H_{\rm env},
\end{equation}
where $H_{\rm sys}$ is the Hamiltonian of the system of interest, $H_{\rm int}=\sum_{\bf k}(g_{\bf k }^*La_{\bf k}^\da+g_{\bf k}L^\da a_{\bf k})$ is the interaction Hamiltonian and $H_{\rm env}=\sum_{\bf k}\omega_{\bf k}a_{\bf k}^\da a_{\bf k}$ describes a quantized field (environment). Note that $L$ is a system operator characterizing the mutual coupling between the system and the environment. At zero temperature $T=0$, the environmental correlation function is determined by the noise operator $B(t)=\sum_{\bf k} g_{\bf k}a_{\bf k}e^{-i\omega_{\bf k}t}$ in the interaction picture: $
\al(t,s)=\langle 0|[B(t)+B^\dag(t)][B(s)+B^\dag(s)]|0\rangle =\sum_{\bf k}|g_{\bf k}|^2e^{-i\omega_{\bf k}(t-s)}$.

Here we use $|\Psi_{\rm tot}(t)\rangle$ to represent the state of the total system at time $t$, then the reduced density operator $\rho_t$ for the system of interest is given by $\rho_t= {\rm Tr}_{\rm env}[|\Psi_{\rm tot}\rangle\langle\Psi_{\rm tot}|]$ obtained by tracing out the environmental degrees of freedom. If the system and its environment are initially uncorrelated, it has been shown that the density operator for the open system can be decomposed into a set of continuous quantum trajectories living in the system's Hilbert space, denoted by $\psi_t(z^*)$. The trajectory $\psi_t(z^*)$ is governed by a linear stochastic Schr\"odinger equation, termed linear QSD equation \cite{Strunz97,Strunz98}:
\begin{equation}\label{LQSD}
\pa_t\psi_t(z^*)=\big(-iH_{\rm sys}+Lz_t^* -
L^\da\int_0^tds\alpha(t,s)\frac{\delta}{\delta z_s^*}\big)\psi_t(z^*),
\end{equation}
where $z_t^*=-i\sum_{\bf k}g_{\bf k}^*z^*_{\bf k}e^{i\omega_{\bf k}t}$ is a complex Gaussian process satisfying $M[z_t]=M[z_t^*z_s^*]=0$, and $M[z_tz_s^*]=\alpha(t,s)$. Here $M[\cdot]$ denotes the ensemble average over the classical noise $z_t^*$. Note that $\alpha(t,s)$ is the correlation function defined above. The quantum trajectory $\psi_t(z^*)$ is designed to recover the density operator of the system by taking ensemble average: $\rho_t=M[|\psi_t(z^*)\ra\la\psi_t(z^*)|]=\int \frac{dz^2}{\pi} e^{-|z|^2}|\psi_t(z^*)\ra\la\psi_t(z^*)|$.

The formal non-Markovian QSD equation (\ref{LQSD}) is a remarkable result since it is completely general irrespective of spectral density of environment and coupling strength. The appearance of functional derivative in Eq.~(\ref{LQSD}) is strongly reminiscent of the convolution kernel appearing in the Nakajima-Zwanzig master equations obtained by projection operator technique \cite{Breuer,ZW}. Thus, for the purpose of practical applications, one can recast the existing functional derivative into a time-local form:
$\delta\psi_t(z^*)/\delta z_s^*=O(t,s,z^*)\psi_t(z^*)$. Combined with the consistency condition $\frac{\delta}{\delta z_s^*}\frac{\partial\psi_t}{\pa t}=\frac{\partial}{\partial t}\frac{\delta\psi_t}{\delta z_s^*}$, this form yields the equation of motion of the O-operator
\cite{Strunz98,Strunz99,Yupra99}:
\begin{equation}\label{CC}
\frac{\partial O}{\partial t}=\left[-iH_{\rm sys}+Lz_t^*-L^\da\bar{O},O\right]
-L^\da\frac{\delta\bar{O}}{\delta z_s^*},
\end{equation}
where $\bar{O}(t,z^*)\equiv\int_0^t ds \al(t,s)O(t,s,z^*)$ and the initial condition $O(s,s,z^*)=L$ is satisfied. Once the O-operator can be explicitly constructed, then the original QSD equation (\ref{LQSD}) takes a desirable time-local form:
\begin{equation} \label{TL}
i \partial_t \psi_t
= \left[H_{\rm sys}+iLz_t^*-iL^\da\bar{O}(t,z^*)\right]\psi_t
\equiv H_{\rm eff}\psi_t.
\end{equation}
Clearly, finding the solutions to the nonlinear operator equation (\ref{CC}) is by no means trivial. It can be shown from the stochastic propagator of the linear QSD equation, formally the O-operator defined in the expression $\delta\psi_t(z^*)/\delta z_s^*=O(t,s,z^*)\psi_t(z^*)$ indeed exist, but determination of its explicit expression is typically a difficult issue. Once the explicit time-local QSD equation is derived, we have shown recently that a time-local master equation may be derived directly from the time-local QSD equation \cite{Chenetal2013}. Such a derivation is applicable to a generic $N$-qubit systems coupled to a common bath where a general Lindblad type of time-local master equations may be obtained \cite{Breuer,ACH}. The difficulty in the expansion process of the memory kernel for the time-local master equation has been transferred to the solution of O-operator through consistency condition~(\ref{CC}). Up to now, several physically interesting examples have been explicitly solved \cite{Strunz98,Strunz99,Jing-Yu10}. Here for the first time, we have derived the time-local QSD equations for a large class of many-qubit systems that are of importance in atomic many-body physics, quantum information science and quantum optics.

For numerical simulations, one must use the normalized pure states
$\ti{\psi}_t(z^*)=\frac{\psi_t(z^*)}{||\psi_t(z^*)||}$ with
$||\psi_t(z^*)||=\langle\psi_t(z^*)|\psi_t(z^*)\rangle^{1/2}$, which is governed by a nonlinear version of the QSD equation \cite{Strunz98}:
\begin{eqnarray*} 
\frac{d}{dt}\ti{\psi}_t&=&-iH_{\rm
sys}\ti{\psi}_t+\Delta_t(L)\ti{z}_t^*\ti{\psi}_t \\
\label{nLQSD} &-&\Delta_t(L^\da)\bar{O}(t,\ti{z}^*)\ti{\psi}_t+\la
\Delta_t(L^\da)\bar{O}(t,\ti{z}^*)\ra_t\ti{\psi}_t,
\end{eqnarray*}
where $\Delta_t(A)\equiv A-\la A\ra_t$ for any operator $A$, $\la
A\ra_t\equiv\la\ti{\psi}_t|A|\ti{\psi}_t\ra$ denotes the quantum average, and $\ti{z}_t^*=z_t^*+\int_0^tds\alpha^*(t,s)\la L^\da\ra_s$ is the shift noise. For all the numerical results to be presented below, we always use the normalized nonlinear QSD equation.

\section{Time-Local Non-Markovian O-operator for $N$-qubit System}\label{theorem}

Equation~(\ref{CC}) manifests that the O-operator may be determined by the commutation relations involving $H_{\rm sys}$ and $L$ together with a set of basis operators. In general, the O-operator may be expanded as \cite{Yupra99},
\begin{equation}\label{Oop}
O(t,s,z^*)=O^{(0)}(t,s)+\sum_{k=1}^M O^{(k)}(t,s,z^*),
\end{equation}
where $O^{(0)}(t,s)=\sum_jf_j(t,s)O^{(0)}_j$ is the noise-free term;
$O^{(1)}(t,s,z^*)=\sum_j\int_0^tp^{(1)}_j(t,s,s_1)z_{s_1}^*ds_1O^{(1)}_j$ is the linear-noise term, and in general
$O^{(k)}(t,s,z^*)=\sum_j\int_0^t\cdots\int_0^t
p^{(k)}_j(t,s,s_1,\cdots,s_k)z_{s_1}^*\cdots z_{s_k}^*ds_1...ds_kO^{(k)}_j$
contains the $k$th-order polynomial noises. The O-operator contains up to $M$th-order noise integral and a finite $M$ means the corresponding model could be solved exactly. Note that all the basis operators $O^{(k)}_j$'s and the functions $f_j$'s, $p^{(k)}_j$'s are noise-free. Moreover, the basis operators $O^{(k)}_j$'s are time-independent. The equations for $f_j$'s and $p^{(k)}_j$'s may be obtained from Eq.~(\ref{CC}). Below, we shall show that convergent and polynomial O-operators can be explicitly constructed for many-qubit models.  \\

{\bf Theorem} For the open system model with
\begin{equation}\label{HL}
H_{\rm sys}=\frac{\om}{2}\sum_{j=1}^N\sigma^{(j)}_z,\quad L=\sum_{j=1}^N\sigma^{(j)}_-,
\end{equation}
the exact O-operator (\ref{Oop}) can be explicitly determined. It contains up to $M=(N-1)$-order noises. That is $O^{(k)}_j=0$ if $k\geqslant N$.

\begin{table}[h!]
\begin{tabular}{|c|c|c|c|c|c|c|c|c|c|} \hline
$N \setminus k$ & $0$ & $1$ & $2$ & $3$ & $4$ & $5$ & $6$ & $7$ & $8$\\ \hline
$1$ & $1$ & $0$ & $0$ & $0$ & $0$ & $0$ & $0$ & $0$ & $0$\\ \hline
$2$ & $2$ & $1$ & $0$ & $0$ & $0$ & $0$ & $0$ & $0$ & $0$\\ \hline
$3$ & $4$ & $2$ & $1$ & $0$ & $0$ & $0$ & $0$ & $0$ & $0$\\ \hline
$4$ & $6$ & $4$ & $2$ & $1$ & $0$ & $0$ & $0$ & $0$ & $0$ \\ \hline
$5$ & $9$ & $6$ & $4$ & $2$ & $1$ & $0$ & $0$ & $0$ & $0$\\ \hline
$6$ & $12$ & $9$ & $6$ & $4$ & $2$ & $1$ & $0$ & $0$ & $0$\\ \hline
$7$ & $16$ & $12$ & $9$ & $6$ & $4$ & $2$ & $1$ & $0$ & $0$ \\ \hline
$8$ & $20$ & $16$ & $12$ & $9$ & $6$ & $4$ & $2$ & $1$ & $0$  \\ \hline
$9$ & $25$ & $20$ & $16$ & $12$ & $9$ & $6$ & $4$ & $2$ & $1$  \\ \hline
\end{tabular}
\caption{This table summarizes the explicit constructions of the O-operator for $N$ qubits system with identical frequency. The notation $k$ stands for the orders of noises contained in the O-operator for the $N$-qubit dissipation model ($N=1,2,\cdots,9)$. For instance, when $N=2$, $m(2,0)=2, m(2,1)=1, m(2,k)=0 \,\, (k \geqslant 2)$.} \label{qutab}
\end{table}
The general proof of {\bf Theorem} for an arbitrary $N$ is rather cumbersome (For details, see Appendix \ref{appenda}). If we use $m(N,k)$ ($0\leqslant k\leqslant N-1$) to denote the number of terms with $k$-fold noise integration in the O-operator, we have the following relations indicated in Table \ref{qutab}:
\begin{eqnarray}
\label{mN0} m(N,0)&=&m(N-2,0)+N, \\
\label{mNk} m(N,k)&=&m(N-1,k-1),
\end{eqnarray}
with $m(1,0)=1$. Table \ref{qutab} lists the numbers $m(N,k)$ for the models up to $9$ qubits. Obviously, when $N=1$, the O-operator just reduces to $O(t,s,z^*)=f(t,s)\sigma_-$ for the case of a single qubit \cite{Strunz98} with $H_{\rm sys}=\frac{\om}{2}\sigma_z$ and $L=\sigma_-$. When $N=2$, it is easy to check that $O(t,s,z^*)=f_1(t,s)O^{(0)}_1+f_2(t,s)O^{(0)}_2
+i\int_0^t ds_1p(t,s,s_1)z_{s_1}^*O^{(1)}_1$, where
$O^{(0)}_1=\sigma_-^A+\sigma_-^B$,
$O^{(0)}_2=\sigma_z^A\sigma_-^B+\sigma_-^A\sigma_z^B$, and
$O^{(1)}_1=\sigma_-^A\sigma_-^B$. As an application, we point out that the exact two-qubit time-local QSD equation allows us to calculate the entanglement evolution of the density matrix constructed from non-Markovian quantum trajectories \cite{Zhaoetal2011}. Similarly, the exact equations for the three-qubit O-operator can be determined explicitly (See Appendix \ref{appendb}).

The result in {\bf Theorem} regarding O-operators can be generalized to the $N$-qubit systems where each qubit has a different frequency $H_{\rm sys}=\sum_j\omega_j\sigma^{(j)}_z/2$ and the general coupling operator $L=\sum_j g_j \sigma^{(j)}_-$. It is expected that more basis operators will be needed in the O-operator construction listed in Table \ref{qutab} for $k\leqslant N-2$. Actually, they could be obtained by decomposing those operators for the isotropic case in Eq.~(\ref{HL}). In the end of Appendix \ref{appendb}, we have given explicitly the basis operators for the three-qubit O-operator. Furthermore, we point out that the non-Markovian QSD can be used to simulate a multiple high-spin dissipative model $H_{\rm sys}=\sum_n\omega_nJ^{(n)}_z$, whose total spin number is $N=\sum_nj_n$. The O-operator also has up to $(N-1)$-th order of noise integral \cite{multilevel}. However, it is easy to see that the O-operators for a N-qubit system are generally very different from the O-operators for the specific N-level system considered in \cite{multilevel}.

\section{Numerical results and discussions}\label{result}

Below, numerical results for the non-Markovian quantum dynamics of many-body systems are presented. For simplicity, and for recovering the Markov limit, we assume that the correlation function of the environment is described by a Ornstein-Uhlenbeck process:
\begin{equation}\label{OU}
\alpha(t,s)=\frac{\gamma}{2}e^{-\gamma|t-s|}.
\end{equation}
Clearly, the Ornstein-Uhlenbeck noise recovers the Markov limit when $\gamma\rightarrow\infty$. It should be noted, however, that our time-local QSD equations are valid and available for arbitrary types of correlation functions.

\begin{figure}[htbp]
\centering
\includegraphics[width=3in]{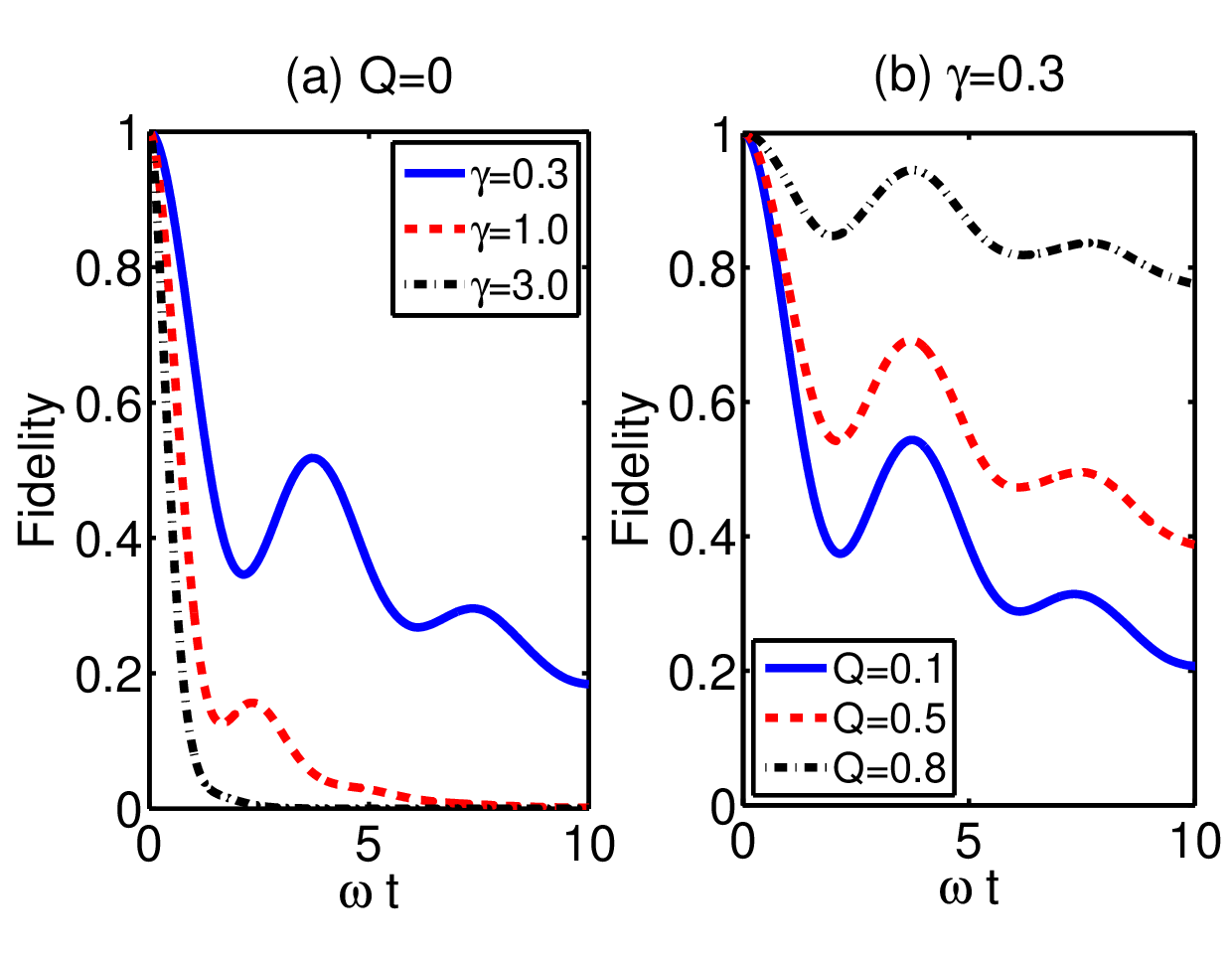}
\caption{(Color online) Evaluating fidelity under the influence of non-Markovian amplitude damping noise via quantum trajectories ($1000$ realizations \cite{comment}). The initial state is the Werner state with parameter $Q$: $W=\frac{Q}{8}I_8+(1-Q)|\psi_0\rangle\langle\psi_0|$, where
$|\psi_0\rangle=(1/\sqrt{3})(|100\rangle+|010\rangle+|001\rangle)$. (a) Fixed initial pure state with $Q=0$ for different $\gamma$; (b) Fidelity for different mixed states for the same noise with $\gamma=0.3$. }
\label{threeq}
\end{figure}

The first example shows how a three-qubit system evolves when coupled to a common multiple-mode environment. The exact time-local QSD equation can be derived explicitly (For details, see Appendix \ref{appendb}). Our numerical simulations with the zeroth-order $O^{(0)}(t,s)$ and the first-order terms $O^{(1)}(t,s,z^*)$ reveal some novel features of coherence dynamics measured by quantum fidelity. Fig.~\ref{threeq}(a) shows the plot of fidelity against time for different environmental memory times $\tau=1/\gamma, \gamma=0.3, 1.0, 3.0$. Clearly, the fidelity is profoundly affected by the memory times. The result suggests a rather interesting feature that quantum coherence can typically survive longer in a non-Markovian dissipative environment. Another interesting feature arising from this system is that, for a strong non-Markovian environment with $\gamma=0.3$, the quantum fidelity is closely related to the degree of entanglement of initial three-qubit Werner states measured by the parameter $Q$ [Fig.~\ref{threeq}(b)]. As the initial state approaches separable state, the state becomes less affected by the environment reflecting the fragile feature of an entangled state \cite{Yu-Eberly08}.

The exact QSD equation Eq.~(\ref{TL}) is known to be a remarkable analytical tool complementing the non-Markovian master equation. In the case of multiple-qubit systems, the exact master equations are still unknown. Here the exact QSD equation is employed in quantum dynamical control of a three-qubit system. Now we consider a three-qubit system, and arrange the order of basis vectors in the following way,
$\{|111\ra,|s_1\ra,|b_1\ra,|c_1\ra,|s_2\ra,|b_2\ra,|c_2\ra,|000\ra\}$, where
$|s_1\ra\equiv(|110\ra+|101\ra+|011\ra)/\sqrt{3}$,
$|a_1\ra\equiv(-2|110\ra+|101\ra+|011\ra)/\sqrt{6}$,
$|b_1\ra\equiv(|101\ra-|011\ra)/\sqrt{2}$,
$|s_2\ra\equiv(|100\ra+|010\ra+|001\ra)/\sqrt{3}$,
$|a_2\ra\equiv(-2|100\ra+|010\ra+|001\ra)/\sqrt{6}$, and
$|b_2\ra\equiv(|010\ra-|001\ra)/\sqrt{2}$,
then the effective Hamiltonian for the exact QSD equation can be written as,
\begin{widetext}
\begin{equation}\label{q3}
H_{\rm eff}=\left(\begin{array}{cccccccc}
3\om/2-3iF_1  & 0 & 0  & 0 & 0  & 0  & 0  & 0 \\
-\sqrt{3}i(2U_1^{(1)}-z_t^*) & \om/2-4iF_4-2iF_3 & 0 & 0 & 0 & 0 & 0 & 0 \\
0 & 0 & \om/2-if & 0 & 0 & 0 & 0 & 0 \\
0 & 0 & 0 & \om/2-if & 0 & 0 & 0 & 0 \\
-\sqrt{3}U_1^{(2)} & -3iU_2^{(1)}+2iz_t^* & 0 & 0 & -\om/2-3iF_2 & 0 & 0 & 0 \\
0 & 0 & iz_t^*/2 & -\sqrt{3}iz_t^*/2 & 0 & -\om/2 & 0 & 0  \\
0 & 0 & -\sqrt{3}iz_t^*/2 & -iz_t^*/2 & 0 & 0 & -\om/2 & 0 \\
0 & 0 & 0 & 0 & 0 & 0 & 0 & -3\om/2
\end{array}\right),
\end{equation}
\end{widetext}
where $f\equiv F_4-F_3$, $U_j^{(1)}\equiv\int_0^tdsP_j^{(1)}(t,s)z_s^*$,
$j=1,2$, $U_1^{(2)}\equiv
\int_0^t\int_0^tds_1ds_2P_1^{(2)}(t,s_1,s_2)z_{s_1}^*z_{s_2}^*$, and all of these functions and coefficients could be found in Appendix \ref{appendb}. From Eq.~(\ref{q3}), we can easily identify a decoherence-free subspace (DFS) spanned by $|b_2\ra$ and $|c_2\ra$, in which entangled states can be protected. In contrast, in the case of two-qubit or  qubit-qutrit systems, the DFS only contains the state $(|10\ra-|01\ra)/\sqrt{2}$.

\begin{figure}[htbp]
\centering
\includegraphics[width=3in]{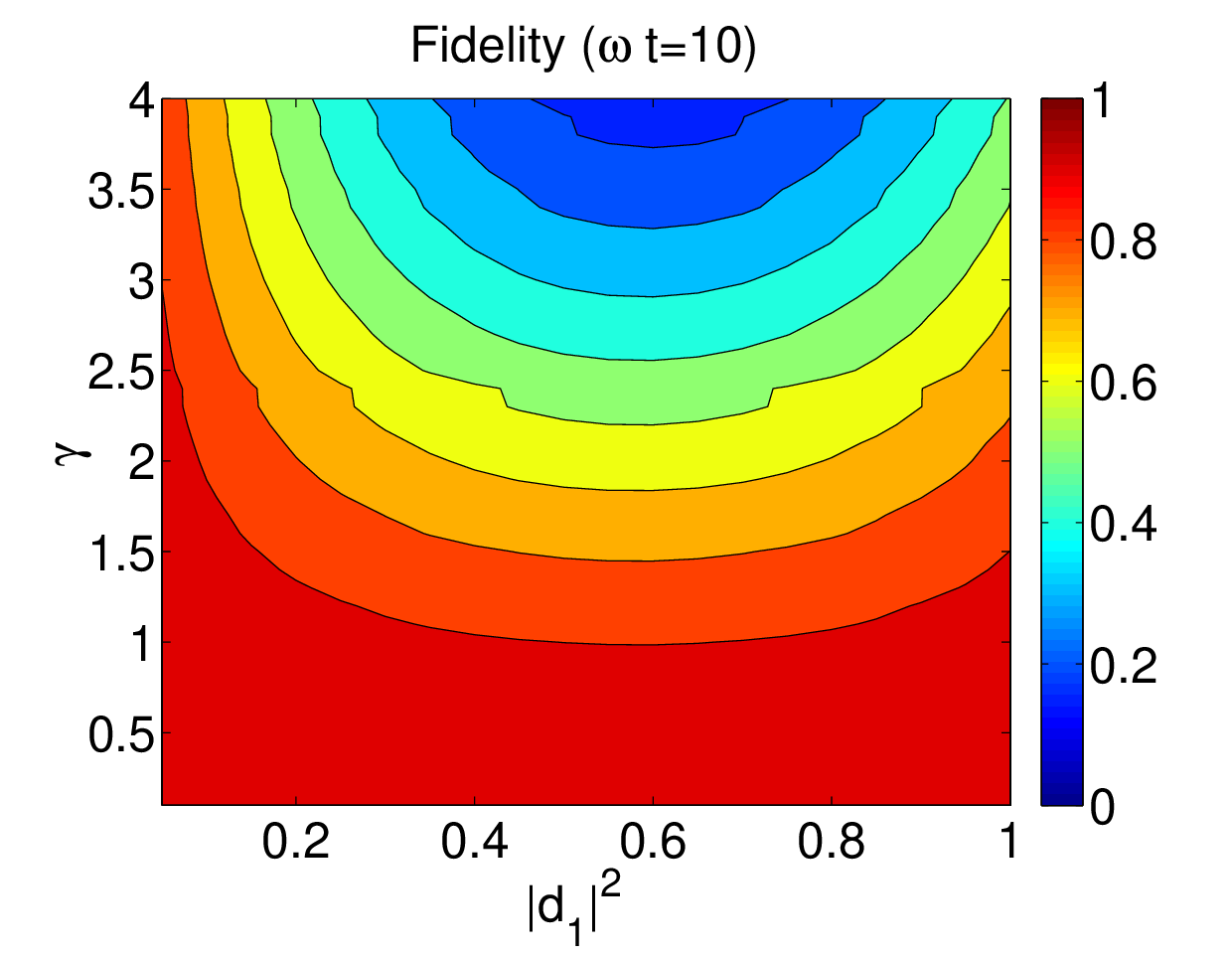}
\caption{(Color online) Fidelity [see Eq.~(\ref{fi})] under the control of a periodical rectangular pulse sequence $c(t)$, whose period, duration time and strength are $T$, $\De$ and $\Phi/\De$, respectively. The initial state is an arbitrary pure state with only one excitation. We choose $\Phi=\om$, $T/\De=2$ and take a snap at the moment $\om t=10$.}
\label{dga3d}
\end{figure}

An arbitrary pure state $\psi_0$ living in the one-exciton subspace may be represented by  $|\psi_0\ra=d_1|s_2\ra+d_2|b_2\ra+d_3|c_2\ra$ with
$|d_1|^2+|d_2|^2+|d_3|^2=1$. By Eqs.~(\ref{TL}) and (\ref{q3}),
$|\psi_t(z^*)\ra=e^{i\om t/2}
(e^{-3\int_0^tdsF_2(s)}d_1|s_2\ra+d_2|b_2\ra+d_3|c_2\ra)$, where
$\pa_tF_2(t)=\ga/2+(-\ga+i\om)F_2+3F_2^2$ and $F_2(0)=0$ [see Eq.~(\ref{F2})]. Thus $\rho_t=M[|\psi_t(z^*)\ra\la\psi_t(z^*)|]=|\psi_t\ra\la\psi_t|$ and the fidelity is
\begin{equation}\label{fi}
\la\psi_0|\rho_t|\psi_0\ra=|1-(1-e^{-3\int_0^tdsF_2(s)})|d_1|^2|^2.
\end{equation}
A simple yet efficient control method is to make the integral in Eq.~(\ref{fi}) as small as possible, thus the state can be stabilized in the initial state (fidelity is close to $1$). For this purpose, we consider the external control field applied to the three-qubit system, that is, we replace  $\om$ with $\om+c(t)$ \cite{JWYY}, where $c(t)$ is a control function that will constantly modulate the frequency of the qubit system. Fig.~(\ref{dga3d}) shows the effect of fidelity control for the parameter $|d_1|^2$, and the environmental memory time $\ga$. It is interesting to see that an effective control of fidelity can be made possible only for a small $\ga$, i.e. long memory time $\tau$. When $|d_1|^2$ approaches $0$ or $1$, the condition imposed on the memory time could be relaxed since $|d_1|$ is close to $0$, it means that the initial state has a large overlap with the DFS, so the state is robust against the influence of noise. On the other hand, if the parameter $d_1$ is close to $1$, the fidelity approaches to $|e^{-6\int_0^tdsF_2(s)}|$, where the integral can be very close to zero in the case that the  control parameters in $C(t)$ are chosen properly. From Eq.~(\ref{q3}), one can apply a similar control scheme to the states spanned by $|b_1\ra$ and $|c_1\ra$. In fact, we only need to control the integral of $f(t)$, where $f(t)$ satisfies $\pa_tf(t)=\ga/2+(-\ga+i\om)f+f^2$ and $f(0)=0$ [see Eqs.~(\ref{F3}) and (\ref{F4})]. It is easy to show that the control scheme is also applicable to the other interesting initial states.

\section{Conclusion and outlook}\label{conclusion}

We have established the exact non-Markovian QSD equations for $N$-qubit systems coupled to a common bosonic environment. We discussed the free and controlled dynamics of a three-qubit system to illustrate the power of the exact QSD equation. The results are easily extendable to interacting qubits and high dimensional systems. Our findings will have many applications in many-body quantum coherence dynamics and quantum information science as illustrated by the examples in this paper. In particular, we expect that the results will be useful for research on many-qubit coherence and entanglement control as shown in a three-qubit model presented in this paper.

In addition, our results can motivate other lines of research. Clearly, it is important to apply the many-body QSD equations to the important non-Markovian physical systems such as atomic ensembles in an optical cavity and atomic dynamics in photonic crystals. Furthermore, it would be of interest to extend the current methods to the case of hybrid quantum systems consisting of both continuous and discrete variables.

\acknowledgments

We thank J. H. Eberly and B. L. Hu for useful discussions. We acknowledge grant support from the NSF PHY-0925174, DOD/AF/AFOSR No. FA9550-12-1-0001, Natural Science Foundation of China Grant Nos.~91121015 and 11304031, the National Basic Research Program of China Grant No.~2014CB921401, and the NSAF Grant No.~U1330201.

\appendix

\section{Proof for the O-operator in N-qubit dissipative model}\label{appenda}

The commutation properties ($[\sigma_z, \sigma_-]=-2\sigma_-$, $[\sigma_+, \sigma_-]=\sigma_z$, $[\sigma_+\sigma_-, \sigma_-]=-\sigma_-$, $\cdots$) in the consistency condition of Eq.~(\ref{CC}) for the system Hamiltonian and coupling operator in Eq.~(\ref{HL}) ensures that we can always find a closed set of basis operator to form the O-operator. Yet one should note that the choice of basis could be arbitrary and sometimes redundant. As long as the O-operator complies with Eq.~(\ref{CC}), they must give rise to a unique solution of the QSD equation as well as the dynamics. Consequently, it is easy to see we may obtain a set of exact integro-differential equations for the O-operator, hence we can establish an exact equation for the dynamics of the multiple-qubit model.

Furthermore, for the case with non-identical transition frequencies of qubits and asymmetrical couplings to the common bath, we can still obtain the exact equations for the O-operator which typically contains more terms than that given in the Table \ref{qutab}.

Now we start to prove Table \ref{qutab} or Eqs.~(\ref{mN0}) and (\ref{mNk}). As the matter of understanding convenience, we define a special ``minus exciton number'' $N_{me}$ for each operator basis (It must be a product of Pauli matrix for every qubit, i.e. $\si_{x1}^{(1)}\si_{x2}^{(2)}\cdots\si_{xN}^{(N)}$, where
$xj\in\{+,-,z,0\}$, $j=1,2,\cdots,N$ and $\si_0\equiv I$) in O-operator. It is settled that $N_{me}=0$ for the identical operator $I$ and $\sigma_z$ and $N_{me}=\mp1$ for $\sigma_\pm$. The number $N_{me}$ for an operator basis is determined by the addition of that for each qubit. Evidently,
$N_{me}(\sigma_+^{1}\sigma_-^{2})=(-1)+1=0$,
$N_{me}(\sigma_-^{1}\sigma_-^{2})=1+1=2$,
$N_{me}(\sigma_z^{1}\sigma_-^{2})=0+1=1$, etc. And the minus exciton number for an arbitrary operator as a combination of operator bases with the same $N_{me}=n$ also equals to $n$. Therefore, $N_{me}(H_{\rm sys})=0$, $N_{me}(L)=1$ and $N_{me}(L^\da L)=0$. When starting to construct an exact O-operator, we need to insert Eq.~(\ref{Oop}) (the first term could be usually chosen as $L$ or a part of $L$ due to the initial condition), into Eq.~(\ref{CC}). Then we encounter with three commuters and one functional derivative. It is easy to find
\begin{eqnarray}\label{Nme1}
N_{me}\left([H_{\rm sys}, O_j^{(k)}]\right)&=&N_{me}(O_j^{(k)})=k+1, \\ \non
N_{me}\left([Lz_t^*,O_j^{(k)}]\right)&=&N_{me}(O_j^{(k)})+1
=N_{me}(O_j^{(k+1)}) \\ \label{Nme2} &=&k+2, \\ \non
N_{me}\left([L^\da O_l^{(k')},O_j^{(k)}]\right)
&=&N_{me}(O_l^{(k')})+N_{me}(O_j^{(k)})-1\\ \label{Nme3} &=&k+k'+1, \\  \non
N_{me}\left(L^\da\frac{\delta\bar{O}^{(k)}(t,z^*)}{\delta z^*_s}\right)
&=&N_{me}(O_j^{(k)})-1=N_{me}(O_j^{(k-1)}) \\ \label{Nme4} &=&k, \quad k\geqslant1.
\end{eqnarray}
We can conclude that (i) the commuter operation $[H_{\rm sys}, \cdot]$ only generate the operator bases with the same order of noise integral; (ii) the commuters $[Lz_t^*, \cdot]$ and $[L^\da\bar{O}, \cdot]$ in Eq.~(\ref{CC}) will bring more and more new operators with larger and larger $N_{me}$ into the construction of O-operator; (iii) the functional derivative could be used to derive the boundary conditions between $O_j^{(k)}$ and $O_j^{(k-1)}$, and it is also consistent with the existence of $O_j^{(0)}$ that must be found in the coupling operator $L$.

To our model in Eq.~(\ref{HL}), the iteration is finite since for an $N$-qubit system, $\max\{N_{me}\}=N$, which corresponds to the operator basis $\prod_{j=1}^N\sigma_-^{(j)}$. We can check $[Lz_t^*,
\prod_{j=1}^N\sigma_-^{(j)}]=0$ and $[L^\da O_l^{(k')},
\prod_{j=1}^N\sigma_-^{(j)}]=0$ with $k'\geqslant1$ by Eqs.~(\ref{Nme2}) and (\ref{Nme3}). Therefore $M=N-1$ and $O^{(M)}$ could be chosen as
$\prod_{j=1}^N\sigma_-^{(j)}$ or its multiplier. Thus $m(N,N-1)=1$, which is independent on the choice of O-operator bases. For the other order of O-operators $O_j^{(k)}$, they must be constructed by operator basis with ``minus exciton number'' $N_{me}=k+1$.

Now we can start to find the basis operator $O_j^{(k)}$, $0\leqslant k\leqslant M$, explicitly. Here we rewrite $O_j^{k}$ into $O_{N,j}^{(k)}$ to indicate there are $N$ qubits in the system. For $N=1$, $O_{1,1}^{(0)}=\sigma_-$; For $N=2$, $O_{2,1}^{(0)}=\sigma_-^{(1)}+\sigma_-^{(2)}$,
$O_{2,2}^{(0)}=\sigma_-^{(1)}\sigma_z^{(2)}+\sigma_-^{(2)}\sigma_z^{(1)}$, $O_{2,1}^{(1)}=\sigma_-^{(1)}\sigma_-^{(2)}$ (These two cases have been justified in the previous works about single qubit and two-qubit dissipative dynamics respectively by QSD method); For $N\geqslant3$, they could be chosen as
\begin{eqnarray}
O_{N,1}^{(0)}&=&\sum_{i=1}^N\sigma_-^{(i)}, \\ \non
O_{N,j}^{(0)}&=&\sum_p\left[\left(\prod_{i=n_{N+1-j}}^{n_{N-1}}\sigma_z^{(i)}
\right)\sigma_-^{(n_N)}\right], \quad N\geqslant
j\geqslant2, \\ \label{ON0} O_{N,j}^{(0)}&=&\sum_p\left(
O_{N-2,j-N}^{(0)}\sigma_+^{(n_{N-1})}\sigma_-^{(n_{N})}\right), \quad
 j>N  \\ \label{ONk}
O_{N,j}^{(k)}&=&\sum_p\left(O_{N-1,j}^{(k-1)}\sigma_-^{(n_N)}\right),
\end{eqnarray}
where $\sum_p$ means the summation of all kinds of permutation of $N$ different numbers $n_1,\cdots,n_N$ over $1,2,\cdots,N$. These bases are not only linear independent but also completed. And for the model with nonidentical qubits, we have to exhaust each term with $N_{me}(O_j^{k})=k+1$, and their number is indeed also finite.

Then the third step is to apply the idea of mathematical induction to verify Eqs.~(\ref{ON0}) and (\ref{ONk}), which are obviously equivalent to Eqs.~(\ref{mN0}) and (\ref{mNk}) respectively. By the construction method given in Eqs.~(\ref{ON0}) and (\ref{ONk}), it is straightforwardly to write the O-operator for the three identical qubits model explicitly:
\begin{eqnarray}\non
O_1^{(0)}&=&\sigma_-^{(1)}+\sigma_-^{(2)}+\sigma_-^{(3)}, \\ \non
O_2^{(0)}&=&(\si_z^{(1)}+\si_z^{(2)})\si_-^{(3)}+(\si_z^{(1)}+\si_z^{(3)})
\si_-^{(2)}\\ \non &+&(\si_z^{(2)}+\si_z^{(3)})\si_-^{(1)}, \\ \non
O_3^{(0)}&=&\si_z^{(1)}\si_z^{(2)}\si_-^{(3)}+\si_z^{(1)}\si_z^{(3)}
\si_-^{(2)}+\si_z^{(2)}\si_z^{(3)}\si_-^{(1)}, \\ \non
O_4^{(0)}&=&\si_-^{(1)}\si_-^{(2)}\si_+^{(3)}+\si_-^{(1)}\si_-^{(3)}
\si_+^{(2)}+\si_-^{(2)}\si_-^{(3)}\si_+^{(1)}, \\ \non
O_1^{(1)}&=&\si_-^{(1)}\si_-^{(3)}+\si_-^{(2)}\si_-^{(3)}
+\si_-^{(1)}\si_-^{(2)}, \\ \non
O_2^{(1)}&=&\si_-^{(1)}\si_-^{(2)}\si_z^{(3)}+\si_-^{(1)}\si_-^{(3)}
\si_z^{(2)}+\si_-^{(2)}\si_-^{(3)}\si_z^{(1)}, \\ \label{3qu}
O_1^{(2)}&=&\si_-^{(1)}\si_-^{(2)}\si_-^{(3)}.
\end{eqnarray}
$O_j^{(0)}$ ($j=1,2,3,4$), $O_j^{(1)}$ ($j=1,2$), and $O_1^{(2)}$ have exhausted all the possibilities of the symmetry combinations according to $N_{me}=1,2,3$, respectively. It is easy to check they satisfy Eqs.~(\ref{Nme1}), (\ref{Nme2}), (\ref{Nme3}), (\ref{Nme4}) and (\ref{CC}). Through the same iteration process, we could also construct the O-operators for the cases with $N\geqslant4$.

As we stated in the main text, in practice, for each group of operators with the same $k$, we could use any linear combinations of them into the construction of O-operator to reduce the computation complexity as long as the new operator is also consisted by a completed basis. Thus the number of solution is infinite.

\section{Three-qubit model}\label{appendb}

It is instructive to work out the O-operator and the following differential equations of its coefficients for $N=3$ in detail (for a two-qubit case, see, \cite{Zhaoetal2011}). In the isotropic condition of Eq.~(\ref{HL}), it has four terms without noise, two terms containing linear noise, and one term containing double integration over noises:
\begin{eqnarray} \non
O&=&\sum_{j=1}^4f_j(t,s)D^{(0)}_j+\sum_{j=1}^2\int_0^tp_j^{(1)}(t,s,s_1)
z^*_{s_1}ds_1D_j^{(1)}\\ \label{Othreequbit} &+&\iint_0^tp_1^{(2)}(t,s,s_1,s_2)z^*_{s_1}z^*_{s_2}ds_1ds_2D_1^{(2)},
\end{eqnarray}
where we chose another group of operator bases indicated by $D$ to simplify the calculation other than those given in Eq.~(\ref{3qu}). It is also valid since this O-operator satisfies Eq.~(\ref{CC}). Explicitly, the operators in Eq.~(\ref{Othreequbit}) are $D_1^{(0)}=(O_1^{(0)}+O_2^{(0)}+O_3^{(0)})/4$, $D_2^{(0)}=(O_1^{(0)}-O_2^{(0)}+O_3^{(0)})/4$, $D_3^{(0)}=O_4^{(0)}$, $D_4^{(0)}=(O_1^{(0)}-O_2^{(0)})/2$, $D_1^{(1)}=(O_1^{(1)}+O_2^{(1)})/2$, $D_2^{(1)}=(O_1^{(1)}-O_2^{(1)})/2$, and $D_1^{(2)}=O_1^{(2)}$.

And then the initial conditions [By $O(s,s,z^*)=L$] are:
\begin{eqnarray*}
f_1(s,s)&=&f_2(s,s)=f_4(s,s)=1,\\
f_3(s,s)&=&p_1^{(1)}(s,s,s_1)=p_2^{(1)}(s,s,s_1) \\&=&p_1^{(2)}(s,s,s_1,s_2)=0.
\end{eqnarray*}

Substituting Eq.~(\ref{Othreequbit}) into Eq.~(\ref{CC}), and comparing all the coefficients belong to the same basis operators, we finally get:
\begin{equation*}
\partial_tf_1(t,s)=i\omega f_1+3F_1f_1-2F_3f_1-4F_4f_1-2P_1^{(1)}(t,s)
\end{equation*}
\begin{equation*}
\partial_tf_2(t,s)=i\omega f_2+3F_2f_2
\end{equation*}
\begin{eqnarray*}
\partial_tf_3(t,s)&=&i\omega f_3-F_2f_3-2F_2f_4+2F_3f_4+2F_4f_3\\ &+& 2F_4f_4-P_2^{(1)}(t,s),
\end{eqnarray*}
\begin{eqnarray*}
\partial_tf_4(t,s)&=&i\omega f_4-F_2f_3-2F_2f_4+F_3f_3+F_3f_4 \\ &+& F_4f_3+3F_4f_4-P_2^{(1)}(t,s),
\end{eqnarray*}
\begin{eqnarray*}
\partial_tp_1^{(1)}(t,s,s_1)&=&2i\omega p_1^{(1)}+3F_1p_1^{(1)}-3F_2p_1^{(1)} \\
&+&2P_1^{(1)}(t,s_1)f_3+4P_1^{(1)}(t,s_1)f_4\\ &-&3P_2^{(1)}(t,s_1)f_1
-2P_1^{(2)}(t,s,s_1)
\end{eqnarray*}
\begin{eqnarray*}
\partial_tp_2^{(1)}(t,s,s_1)&=&2i\omega p_2^{(1)}+2F_3p_2^{(1)}+4F_4p_2^{(1)} \\ \non
&+&3P_2^{(1)}(t,s_1)f_2
\end{eqnarray*}
\begin{eqnarray*}
\partial_tp_1^{(2)}(t,s,s_1,s_2)&=&3i\omega p_1^{(2)}
+3P_1^{(2)}(t,s_1,s_2)f_2\\ &+& 3F_1p_1^{(2)}+6P_1^{(1)}(t,s_1)p_2^{(1)}(t,s,s_2)
\end{eqnarray*}
together with the boundary conditions:
\begin{eqnarray*}
p_1^{(1)}(t,s,t)&=&2f_1(t,s)-f_3(t,s)-2f_4(t,s), \\
p_2^{(1)}(t,s,t)&=&-2f_2(t,s)+f_3(t,s)+2f_4(t,s),\\
2p_1^{(2)}(t,s,t,s_1)&=&3p_1^{(1)}(t,s,s_1)-3p_2^{(1)}(t,s,s_1).
\end{eqnarray*}

If the correlation function is taken as Ornstein-Uhlenbeck process
in Eq.~(\ref{OU}), and by definitions
$\bar{P}_1^{(1)}(t)\equiv\int_0^t ds\alpha(t,s)P_1^{(1)}(t,s)$,
$\bar{P}_2^{(1)}(t)\equiv\int_0^t ds\alpha(t,s)P_2^{(1)}(t,s)$,
$\bar{P}_1^{(2)}(t,s_1)\equiv\int_0^t ds\alpha(t,s)P_1^{(2)}(t,s,s_1)$, and
$\tilde{P}_1^{(2)}(t)\equiv\int_0^t ds\alpha(t,s)\bar{P}_1^{(2)}(t,s)$, then after a straightforward derivation, we have:
\begin{eqnarray}\non
\partial_tF_1(t)&=&\frac{\gamma}{2}
+(-\gamma+i\omega)F_1+3F_1^2-2F_1F_3\\ \label{F1}&-& 4F_1F_4-2\bar{P}_1^{(1)}\\ \label{F2}
\partial_tF_2(t)&=&\frac{\gamma}{2}+(-\gamma+i\omega)F_2+3F_2^2\\ \non
\partial_tF_3(t)&=&(-\gamma+i\omega)F_3-F_2F_3-2F_2F_4+4F_3F_4 \\ \label{F3}
&+&2F_4^2-\bar{P}_2^{(1)} \\ \non
\partial_tF_4(t)&=&\frac{\gamma}{2}+
(-\gamma+i\omega)F_4-F_2F_3-2F_2F_4+F_3^2 \\ \label{F4} &+&2F_3F_4+3F_4^2-\bar{P}_2^{(1)}
\end{eqnarray}
\begin{eqnarray}\non
\partial_t\bar{P}_1^{(1)}(t)&=&(-2\gamma+2i\omega)\bar{P}_1^{(1)}
+\frac{\gamma}{2}(2F_1-F_3-2F_4)\\ \non &+&(3F_1-3F_2
+2F_3+4F_4)\bar{P}_1^{(1)}\\ &-&3F_1\bar{P}_2^{(1)}-2\ti{P}_1^{(2)} \\ \non
\partial_t\bar{P}_2^{(1)}(t)&=&(-2\gamma+2i\omega)\bar{P}_2^{(1)}
+\frac{\gamma}{2}(-2F_2+F_3+2F_4) \\
&+&2F_3\bar{P}_2^{(1)}+4F_4\bar{P}_2^{(1)}+3F_2\bar{P}_2^{(1)}
\end{eqnarray}
\begin{eqnarray}\non
\partial_t\ti{P}_1^{(2)}(t)&=&(-3\gamma+3i\omega)\ti{P}_2^{(1)}
+\frac{\gamma}{4}(3\bar{P}_1^{(1)}-3\bar{P}_2^{(1)})\\ &+&3F_1\ti{P}_1^{(2)}+3F_2\ti{P}_1^{(2)}
+6\bar{P}_1^{(1)}\bar{P}_2^{(1)}
\end{eqnarray}

In the anisotropic case where $H_{\rm sys}=\frac{\om_1}{2}\si_z^{(1)}
+\frac{\om_2}{2}\si_z^{(2)}+\frac{\om_3}{2}\si_z^{(3)}$, and $L=g_1\si_-^{(1)}+g_2\si_-^{(2)}+g_3\si_-^{(3)}$, we can show that, although all of the symmetries are broken, one can still obtain the basis operators for this general three qubit model. The most straightforward way is to decompose those operators in Eq.~(\ref{3qu}), i.e. to take each single product term as one individual basis operators. All together, it will have $22$ terms:
\begin{eqnarray}\non
&&O_1^{(0)}=\sigma_-^{(1)}, O_2^{(0)}=\sigma_-^{(2)}, O_3^{(0)}=\sigma_-^{(3)}, \\ \non
&&O_4^{(0)}=\si_z^{(1)}\si_-^{(3)}, O_5^{(0)}=\si_z^{(2)}\si_-^{(3)}, \\ \non
&&O_6^{(0)}=\si_z^{(1)}\si_-^{(2)}, O_7^{(0)}=\si_z^{(3)}\si_-^{(2)}, \\ \non &&
O_8^{(0)}=\si_z^{(2)}\si_-^{(1)}, O_9^{(0)}=\si_z^{(3)}\si_-^{(1)}, \\ \non
&&
O_{10}^{(0)}=\si_z^{(1)}\si_z^{(2)}\si_-^{(3)},
O_{11}^{(0)}=\si_z^{(1)}\si_z^{(3)}\si_-^{(2)},  \\ \non &&
O_{12}^{(0)}=\si_z^{(2)}\si_z^{(3)}\si_-^{(1)},
O_{13}^{(0)}=\si_-^{(1)}\si_-^{(2)}\si_+^{(3)}, \\ \non &&
O_{14}^{(0)}=\si_-^{(1)}\si_-^{(3)}\si_+^{(2)},
O_{15}^{(0)}=\si_-^{(2)}\si_-^{(3)}\si_+^{(1)}, \\ \non &&
O_1^{(1)}=\si_-^{(1)}\si_-^{(3)}, O_2^{(1)}=\si_-^{(2)}\si_-^{(3)}, \\ \non &&
O_3^{(1)}=\si_-^{(1)}\si_-^{(2)},O_4^{(1)}=\si_-^{(1)}\si_-^{(2)}\si_z^{(3)},
\\ \non && O_5^{(1)}=\si_-^{(1)}\si_-^{(3)}\si_z^{(2)},
O_6^{(1)}=\si_-^{(2)}\si_-^{(3)}\si_z^{(1)}, \\  &&
O_1^{(2)}=\si_-^{(1)}\si_-^{(2)}\si_-^{(3)}.
\end{eqnarray}
So in the O-operator construction, we have $15$ noise-free terms, $6$ linear-noise terms, and $1$ double-fold-noise term.


\begin{thebibliography}{99}
\bibitem{comment}
We should point out that the number of realization needed to achieve an accurate result varies depending on initial states, correlation functions etc.
\end{thebibliography}
\end{document}